\documentclass{ar-1col-S2O}
\usepackage[numbers]{natbib}
\usepackage{url}
\usepackage{amssymb,amsmath}

\jname{Annu. Rev. Nucl. Part. Sci.}
\jvol{75}
\jyear{2025}
\doi{10.1146/annurev-nucl-121423-100950}

\begin{document}

\markboth{Altmannshofer and Greljo}{Recent Progress in 
Flavor Model Building}

\title{Recent Progress in 
Flavor Model Building}

\author{Wolfgang Altmannshofer$^1$ and Admir Greljo$^2$
\affil{$^1$Department of Physics, University of California, Santa Cruz, and Santa Cruz Institute for Particle Physics, Santa Cruz, CA 95064, USA; \\ email: waltmann@ucsc.edu}
\affil{$^2$Department of Physics, University of Basel, Klingelbergstrasse 82, CH 4056 Basel, Switzerland; email: admir.greljo@unibas.ch}}

\begin{abstract}
The flavor puzzles remain among the most compelling open questions in particle physics. The striking hierarchies observed in the masses and mixing of charged fermions define the Standard Model (SM) flavor puzzle, a profound structural enigma pointing to physics beyond the SM. Simultaneously, the absence of deviations from SM predictions in precision measurements of flavor-changing neutral currents imposes severe constraints on new physics at the TeV scale, giving rise to the new physics flavor puzzle. This review article provides an overview of a selection of recent advancements in flavor model building, with a particular focus on attempts to address one or both of these puzzles within the quark sector.
\end{abstract}

\begin{keywords}
quarks, leptons, flavor puzzles, beyond the SM, flavor physics 
\end{keywords}
\maketitle

\tableofcontents





\section{INTRODUCTION} \label{sec:intro}

Within the landscape of quantum field theories, the Standard Model (SM) of particle physics is a prominent example of a chiral, spontaneously broken non-abelian gauge theory, wherein anomaly cancellation imposes stringent constraints on the charge assignments of matter fields within a single generation. However, nature often surprises us with unexpected complexity, exemplified by the existence of three generations of particles---an apparent redundancy that has no obvious reason. This replication of fermion gauge representations is referred to as \textit{flavor}. Understanding the origin of flavor remains a longstanding and unresolved question in particle physics.

Flavor physics is presently a highly dynamic field of research, driven by a substantial ongoing experimental program and supported by a series of planned future experiments.
Currently, bottom, kaon, charm, tau, and muon factories operate at full capacity, offering a unique opportunity to deepen our understanding of the SM and uncover new physics. As the Large Hadron Collider (LHC) achieves its design collision energy and transitions into the high-luminosity era~\cite{Cerri:2018ypt}, the emphasis is shifting towards precision measurements. Looking ahead, the prospect of an electron-positron circular collider emerges as the next frontier~\cite{FCC:2018evy, CEPCStudyGroup:2018ghi}. In this era of precision, testing flavor and CP violation takes center stage, offering the potential to discover new physics. This could involve either directly detecting light new physics or observing indirect effects of heavy new physics, which may reveal themselves through rare phenomena such as flavor-changing neutral currents (FCNC), and electric dipole moments (EDMs). These rare and forbidden phenomena are particularly sensitive to new physics at energy scales far beyond the TeV range, which are out of reach for direct searches.

The ongoing, robust experimental program has revitalized interest in advancing theoretical approaches in flavor model building. The goal of flavor model building is twofold. First, it aims to explain the observed hierarchical patterns of fermion masses and mixing angles, known as the SM flavor puzzle (section~\ref{sec:SM_puzzle}). Second, if new physics exists just beyond the electroweak scale, flavor model building seeks to explain why no signs of this new physics have been detected in flavor-changing transitions, known as the new physics (NP) flavor puzzle (section~\ref{sec:NP_puzzle}). This review will explore approaches and models that address both flavor puzzles, evaluating their successes and limitations in the context of current and future experimental data. Our emphasis is on the quark sector, but we will also comment on the flavor of charged leptons and neutrinos where appropriate.
After a comprehensive review of traditional solutions to the flavor puzzles in section~\ref{sec:solutions}, the discussion will shift towards an exploration of recent advancements in the field, which will be the central focus of section~\ref{sec:recent}. 

\section{FLAVOR PUZZLES} \label{sec:puzzles}

In this section, we define and discuss the SM and NP flavor puzzles separately.

\subsection{The Standard Model Flavor Puzzle} \label{sec:SM_puzzle}

The observed masses of quarks and leptons, along with their mixing under the weak nuclear force, reveal a peculiar pattern. Up quarks, down quarks, and charged leptons exhibit a distinct generational mass hierarchy, with each generation differing by approximately two orders of magnitude. Additionally, the Cabibbo-Kobayashi-Maskawa (CKM) quark mixing matrix is nearly a unit matrix, with hierarchically suppressed mixing between generations: $ 1 \gg |V_{12}|\sim \lambda \gg |V_{23}|\sim \lambda^2 \gg |V_{13}| \sim \lambda^3$, with $\lambda \simeq 0.2$. The SM fails to explain the observed hierarchies, a deficiency referred to as \textit{the SM flavor puzzle}.

\begin{figure}[tb]
\centering
\includegraphics[width=1.0\linewidth]{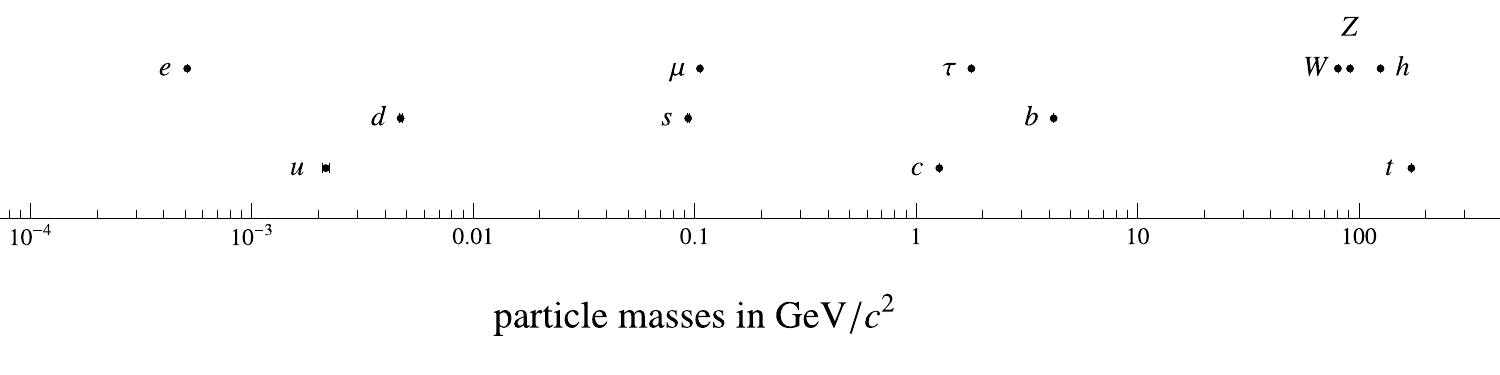} \\[16pt]
\includegraphics[width=1.0\linewidth]{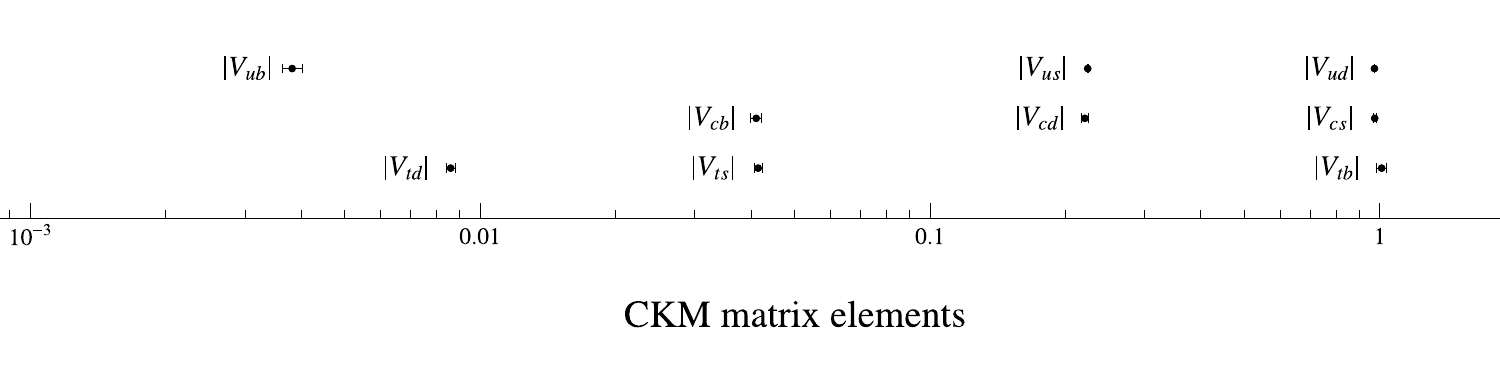}
\caption{Top: Masses of Standard Model particles in units of GeV$/c^2$, including uncertainties. Values for the Higgs boson, gauge bosons, quarks, and leptons are taken from the PDG~\cite{ParticleDataGroup:2024cfk}. Neutrino masses, several orders of magnitude lighter, are not shown. Two neutrino mass differences have been measured \cite{Esteban:2020cvm}, and cosmological data constrain the sum of neutrino masses to $\sum m_\nu \lesssim 0.1$\,eV \cite{Planck:2018vyg}. The lightest neutrino could potentially be massless. Bottom: Absolute values of CKM matrix elements with uncertainties, derived from direct measurements as compiled by the PDG~\cite{ParticleDataGroup:2024cfk}. Figure adapted from Reference~\cite{Altmannshofer:2024ykf}, with permission from the author.}
\label{fig:masses_CKM}
\end{figure}

The origin of masses and mixing in the SM is tied to the renormalizable interactions with a single Higgs field $H$, 
\begin{equation} \label{eq:Yukawa}
    \mathcal{L} \supset -Y_u^{pr} \, \overline{q}_p \widetilde{H} u_r -Y_d^{pr} \, \overline{q}_p H d_r -Y_e^{pr} \, \overline{\ell}_p H e_r\,,
\end{equation}
where $q_p, u_r, d_r, \ell_p$, and $e_r$ are the five gauge representations of the chiral fermions, with $p,r=1,2,3$ as flavor indices and $\widetilde{H} = i \sigma_2 H^*$. The core of the flavor puzzle lies in the following question: Why is there a hierarchical structure in the input parameters $Y_f^{pr}$ when they all enter the theory in the same way, coupling to a single Higgs field? According to the singular value decomposition theorem, $Y_f$ equals $L_f \hat{Y}_f R_f^\dagger$, where $\hat{Y}_f$ is a diagonal matrix with real and positive entries, and $L_f$ and $R_f$ are unitary matrices. The masses are determined by the singular values $\hat{M}_f = \hat{Y}_f \langle H\rangle$, while the quark mixing is given by the CKM matrix $V = L^\dagger_u L_d$. The experimentally determined values of the fermion masses and the CKM matrix elements, shown in Figure~\ref{fig:masses_CKM}, reveal a highly hierarchical structure in the Yukawa matrices. 

It is important to note that the small values of the Yukawa couplings are 'technically natural' in the sense of t'Hooft's naturalness criterion~\cite{tHooft:1979rat}. Setting the Yukawa couplings to zero enhances the symmetry of the theory, which ensures that radiative corrections to these parameters are only logarithmically sensitive to the cutoff scale as long as no new significant sources of flavor violation beyond the SM are introduced. This is in stark contrast to the Higgs hierarchy problem, where the sensitivity to the cutoff scale is quadratic, motivating a symmetry-based solution not far above the electroweak scale. Unfortunately, there is no clear indication of a scale where the flavor puzzle is solved. 

While the small Yukawa couplings are technically natural, they fall short of what one might expect from Dirac's naturalness, where dimensionless input parameters are of order one. Even more perplexing is the specific hierarchical structure of these parameters. We stress that the three $3\times 3$ Yukawa matrices, $Y_u$, $Y_d$, and $Y_e$, are entirely independent of each other. Yet, they all show a similar hierarchical structure in their singular values and an approximate alignment between $Y_u$ and $Y_d$, as reflected by the CKM matrix. This peculiar structure shouts for an explanation.

Compared to quarks and charged leptons, the neutrino sector behaves quite differently, adding further complexity to the puzzle. First, neutrinos have extremely small masses, at least six orders of magnitude smaller than the electron mass. Second, neutrino oscillation experiments reveal large mixing angles~\cite{Esteban:2020cvm}, unlike the small and hierarchical mixing between quarks. Additionally, the ratio of the two observed mass splittings in the neutrino sector is not particularly large when compared to the pronounced mass hierarchies in the charged fermion sector.

The renormalizable SM predicts that neutrinos are massless, calling for physics beyond the SM to explain non-zero neutrino masses. However, when viewing the SM as an effective field theory, the significant mass difference between neutrinos and charged fermions can be explained by the leading higher-dimensional operator~\cite{Weinberg:1979sa}
\begin{equation} \label{eq:Weinberg}
\mathcal L \supset \frac{Y_\nu^{p r}}{\Lambda} \ell_p \ell_r H H ~.
\end{equation}
Here, \(Y_\nu^{pr}\) is a complex-symmetric matrix, diagonalized by \(Y_\nu = U^T_\nu \hat{Y}_\nu U_\nu\), while \(\Lambda\) represents a high-energy scale.
The operator breaks the lepton number by two units, and neutrinos are Majorana particles in this setup. A large energy scale $\Lambda$ where the lepton number is broken provides a compelling explanation for the smallness of the neutrino masses. The see-saw mechanism with heavy right-handed Majorana neutrinos is arguably the simplest realization of this scenario. The Pontecorvo-Maki-Nakagawa-Sakata (PMNS) matrix, the leptonic counterpart to the CKM matrix, exhibits large and seemingly anarchic mixing,\footnote{Interestingly, the PMNS matrix is approximately of the ``tri-bimaximal'' form~\cite{Harrison:2002er} motivating discrete flavor symmetries acting in the lepton sector, see e.g.~\cite{Altarelli:2010gt} for a review.} as one might expect if no special structure is imposed on $Y_\nu^{pr}$. This adds another layer of complexity to the SM flavor puzzle, given that the CKM matrix behaves in a contrasting manner. 

As an alternative to the neutrino mass origin outlined above, the lepton number could remain unbroken, making neutrinos Dirac particles.
In this scenario, their masses would arise from renormalizable Yukawa interactions, similar to the charged fermions in Eq.~(\ref{eq:Yukawa}). The small neutrino masses would then result from extremely tiny Yukawa couplings, further complicating the flavor puzzle. 

Flavor model building attempts to explain the non-generic flavor structure of the SM fermions described above. 
It remains an open question to what extent the flavor hierarchies in the quark sector, the hierarchies in the charged lepton spectrum, and the absence of pronounced hierarchies in the neutrino sector are connected. While this review primarily focuses on the quark sector, we will also discuss the lepton and neutrino sectors where relevant.

\subsection{The New Physics Flavor Puzzle} \label{sec:NP_puzzle}

One practical implication of the SM flavor puzzle, even without knowing its solution, is the existence of approximate accidental symmetries within the SM. The smallness of Yukawa couplings and the CKM alignment leads to approximate flavor and CP conservation, resulting in highly suppressed FCNCs and EDMs. The resulting selection rules, exemplified by the Glashow–Iliopoulos–Maiani (GIM) mechanism~\cite{Glashow:1970gm}, have been instrumental in designing precision tests of the SM that are highly sensitive to new physics, which typically breaks these approximate symmetries. So far, these tests have yielded negative results. However, the null findings have profound consequences for the flavor structure of new physics if it exists near the TeV scale---a situation known as the \textit{new physics flavor puzzle}. That is to say, new flavor parameters of a TeV-scale extension of the SM are also not generic.

To illustrate this point, we employ the Standard Model Effective Field Theory (SMEFT) as a framework to represent short-distance new physics effects~\cite{Grzadkowski:2010es}. By augmenting the SM Lagrangian with a tower of higher-dimensional local operators,
$\mathcal{L} \mathrel{+}= \sum_{\mathcal{O}} C_\mathcal{O}\,\mathcal{O}$,
we introduce new sources of flavor and CP violation that are absent at the renormalizable level. Specifically, we focus on leading-order dimension-6 operators that preserve baryon and lepton numbers. The Wilson coefficients of these operators are parameterized as $ C_{\mathcal{O}} = c_{\mathcal{O}} \Lambda^{-2} $, where $ \Lambda $ is the scale of a new physics completion.

\begin{figure}[tb] \centering
\includegraphics[width=1.0\textwidth]{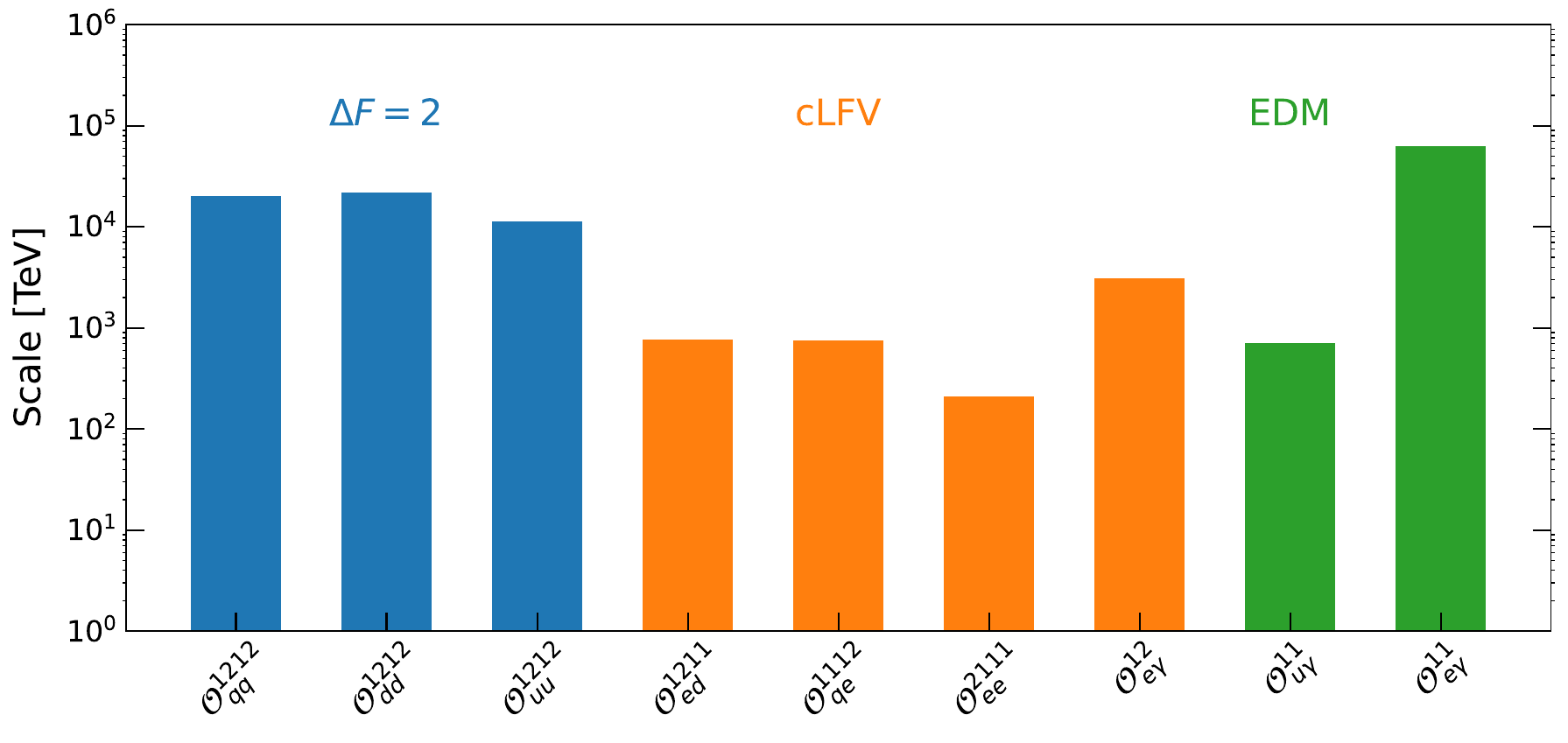}
  \caption{Experimental bounds on a selection of dimension-6 SMEFT operators assuming flavor anarchy. Four-quark operators are considered to be generated at the tree level. Dipole operators contain a loop suppression factor.}
  \label{fig:bounds}
\end{figure}

Consider a flavor-anarchic scenario in which all dimensionless flavor parameters are of order one, assuming no particular hierarchy or structure in the flavor sector of the new physics. For a tree level and one-loop matching, this predicts $c_\mathcal{O} \sim 1$ and  $\sim 1/16\pi^2$, respectively. Under this assumption, many of the dimension-6 operators contribute significantly to processes involving neutral meson oscillations ($ \Delta F = 2 $), charged lepton flavor violation (cLFV), EDMs, or strangeness-changing decays ($ \Delta S = 1 $). The experimental absence of deviations from SM predictions in such processes imposes stringent constraints on the scale $\Lambda$, often pushing it to values as high as six orders of magnitude above the EW scale. For a concrete illustration, Figure~\ref{fig:bounds} presents bounds on a selected set of operators in the Warsaw basis~\cite{Grzadkowski:2010es}, explicitly showing the flavor indices in the down-quark charged-lepton mass basis. For the dipole operators, we include a one-loop suppression factor expected from the matching, whereas the four-quark operators are considered to be generated at the tree level. For more operator examples, see Ref.~\cite{Antusch:2023shi}.

Let us stress that this creates a flavor puzzle only if we anticipate NP to emerge near the TeV scale. Otherwise, a high cutoff scale $\Lambda$ for the SM automatically resolves the tension with experimental data. 

The main motivation for expecting new physics at the TeV scale is the Higgs hierarchy problem, which asks why the Higgs boson mass is so much lighter than the Planck scale despite quantum corrections that should drive it higher. Symmetry-based solutions to this problem---such as supersymmetry (SUSY) or composite Higgs models---imply that new physics to stabilize the Higgs mass is expected not far above the electroweak scale. These theories often introduce new particles and interactions with order-one couplings to the Higgs field. Consequently, the flavor structure of such new physics cannot be generic; it must be hierarchical and (or) aligned to avoid excessive violation of the SM's approximate flavor and CP symmetries. This raises the possibility that the NP and the SM flavor structure might share a common origin.

Secondly, from a purely experimental, bottom-up perspective, the TeV scale is directly accessible to the operating LHC. Direct searches at this energy scale require some couplings to be sizable to produce observable signals, bringing the flavor structure of new interactions into sharp focus. For prompt signals in direct searches, if the new couplings to fermions are not hierarchical or aligned, they could induce dangerous flavor-changing processes. Therefore, understanding the flavor hierarchy in new physics is crucial for a consistent program of direct searches at colliders.

\section{OVERVIEW OF PROPOSED SOLUTIONS} \label{sec:solutions}

The distinctive flavor structure observed in the SM and similarly non-generic flavor patterns in the couplings of TeV-scale new physics demand an explanation. This section provides a concise overview of the most prominent approaches developed to address the two puzzles.

\subsection{Model Building for the Standard Model Flavor Hierarchies}

The observed hierarchies in the quark and charged lepton masses and the CKM mixing angles suggest an underlying mechanism beyond the SM. Here, we explore the two main approaches that have been proposed to address this flavor puzzle: symmetry-based models and geometry-based models.

\subsubsection{Symmetry-Based Models} \label{sec:symmetry}

Symmetry-based models introduce additional symmetries, known as \textit{horizontal} or family symmetries, that act between different generations of fermions. These symmetries constrain the Yukawa couplings, naturally leading to hierarchical mass and mixing patterns after symmetry breaking by small spurions. Depending on the nature of the symmetry group, these models are classified into abelian and non-abelian symmetries. They could be either gauged (fundamental) or global (accidental). The latter can be approximate or (perturbatively) exact. The specifics of the flavor-symmetry-driven mechanism often depend on other aspects of the BSM extension, such as the presence of additional quark-lepton unification or the inclusion of supersymmetry.

The \textit{Froggatt-Nielsen} (FN) mechanism is a seminal example utilizing an \textit{abelian} horizontal symmetry, typically a $U(1)$ symmetry, to explain flavor hierarchies~\cite{Froggatt:1978nt, Leurer:1992wg, Leurer:1993gy}. In this framework, the SM fermions carry generation-dependent charges under the additional $U(1)$ symmetry. A scalar field, known as the FN (or flavon) field ($\Phi$), acquires a vacuum expectation value (vev) $\langle \Phi \rangle$, thus spontaneously breaking the flavor symmetry. The vev $\langle \Phi \rangle$ is assumed to be smaller than the cutoff scale $\Lambda$ of the effective theory. Yukawa couplings arise from higher-dimensional operators suppressed by powers of the small parameter $\varepsilon = \langle \Phi \rangle/\Lambda \ll 1 $. By assigning appropriate $U(1)$ charges, the FN mechanism can qualitatively reproduce the observed mass hierarchies and mixing angles from a single expansion parameter $\varepsilon$. For instance, heavier fermions are assigned lower charges, resulting in less suppression in their Yukawa couplings. In essence, the Yukawa couplings in equation~\eqref{eq:Yukawa} are replaced by the following terms
\begin{equation}
\mathcal L \supset - \tilde Y_u^{pr} \varepsilon^{Q_{q_p}-Q_{u_r}}  \, \overline{q}_p \widetilde{H} u_r -\tilde Y_d^{pr} \varepsilon^{Q_{q_p}-Q_{d_r}} \, \overline{q}_p H d_r -\tilde Y_e^{pr} \varepsilon^{Q_{\ell_p}-Q_{e_r}} \, \overline{\ell}_p H e_r\,,
\end{equation}
where the $\tilde Y_{u,d,e}$ are ``proto'' Yukawa couplings that can all be of $\mathcal O(1)$, and $Q_{q_p}$, $Q_{\ell_p}$, $Q_{u_r}$, $Q_{d_r}$, $Q_{e_r}$ are the $U(1)$ charges of the quarks and leptons. We assume that the Higgs is uncharged $Q_H = 0$, and without loss of generality we have set $Q_\Phi = 1$.

\begin{figure}[tb] \centering
  \includegraphics[width=.8\textwidth]{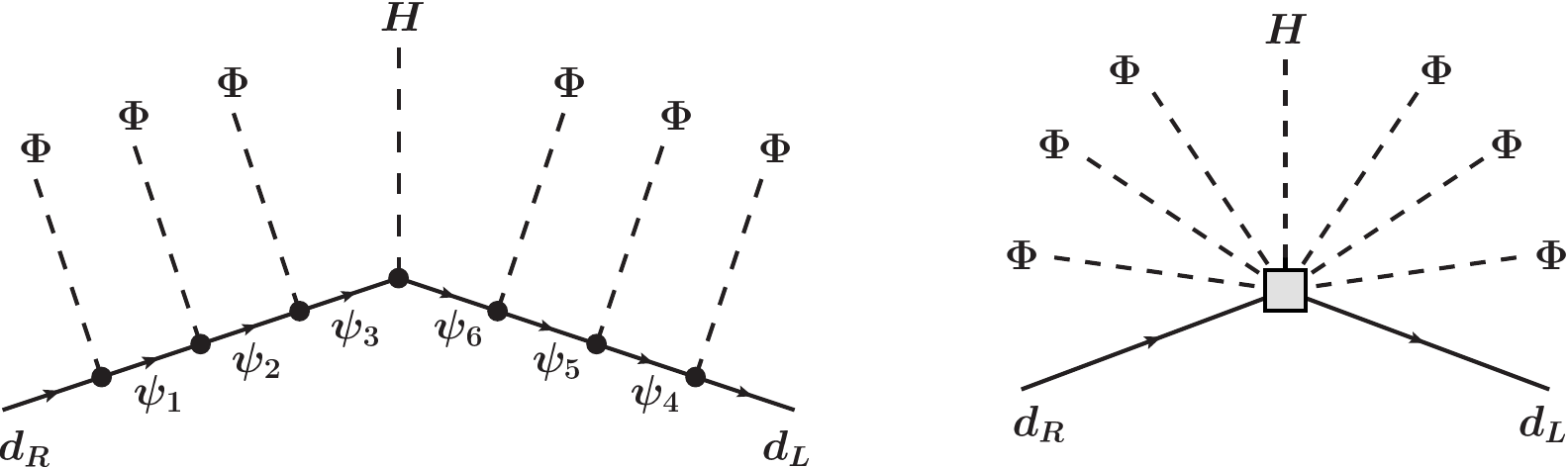}
  \caption{Example of how the Froggatt-Nielsen mechanism might generate a small Yukawa coupling of the down quark. The shown scenario corresponds to the $U(1)$ charges $Q_{q_1} = +3$, $Q_{d_1} = -3$, $Q_{H}= 0$, and $Q_{\Phi}= +1$. Left: Chain of vectorlike fermions $\psi_i$ that connect the down quarks to the Higgs boson. Right: the corresponding higher dimensional operator. Figure adapted from Reference~\cite{Altmannshofer:2024ykf}, with permission from the author.}
  \label{fig:FN}
\end{figure}

The most straightforward UV completion of the higher dimensional operators involves chains of vector-like fermions that communicate the symmetry breaking to the SM fermions. In such a case, the scale $\Lambda$ corresponds to the masses of the vector-like fermions, and the proto Yukawas are combinations of their couplings. A typical example is shown in Figure~\ref{fig:FN}. The FN setup contains a number of free $\mathcal O(1)$ parameters sufficient to reproduce the observed fermion masses precisely. If the $U(1)$ symmetry is gauged, the models contain a $Z^\prime$ gauge boson with a mass of order $\langle \Phi \rangle$. If the $U(1)$ is global, one instead expects a very light pseudo-Nambu-Goldstone boson. These particles generically have flavor-changing couplings and can be effectively searched for with flavor probes. Flavor models commonly feature a decoupling limit: by sending both $\langle \Phi \rangle$  and $\Lambda$ to high energies while keeping $\varepsilon = \langle \Phi \rangle / \Lambda$ constant, these models can address the SM flavor puzzle without introducing excessive FCNCs at low energies. Current experimental constraints impose a lower bound on the scale of flavor dynamics, which are, however, model-dependent. In the simplest cases with gauged $U(1)$ symmetries, the $Z^\prime$ can contribute at the tree level to kaon mixing, and bounds can be as stringent as $\langle \Phi \rangle \gtrsim 10^4$\,TeV. In the case of a global $U(1)$, one often finds even stronger bounds on the new physics scale from flavor-violating decays of kaons and muons into the light pseudo-Nambu-Goldstone boson, see section~\ref{sec:axions} for more details. Interestingly, for a discrete FN, the breaking scale can be as low as $\langle \Phi \rangle \sim$\,TeV; see~\cite{Greljo:2024evt}. Future experiments could potentially reveal evidence for low-scale flavor models.

\bigskip
\textit{Non-abelian flavor symmetries}, such as $SU(3)$ or $SU(2)$, organize the fermion generations into non-trivial group representations. By assigning fermions to multiplets, these symmetries constrain the structure of the Yukawa couplings. The observed flavor structure typically arises through sequential and hierarchical symmetry breaking. Non-abelian symmetries often lead to more restrictive textures than FN models. However, challenges to reproducing realistic parameters
in these models include the complexity of the symmetry-breaking sector.
Models based on $SU(3)$ flavor symmetry have been proposed to generate realistic mass hierarchies; see e.g.~\cite{King:2001uz, Antusch:2008jf, Nardi:2011st, Alonso:2011yg}. However, the sizable Yukawa coupling of the top quark ($y_t \sim 1$) breaks the $SU(3)$ symmetry, making $SU(2)$ (or better $U(2)$) groups more attractive~\cite{Pomarol:1995xc, Barbieri:1995uv, Barbieri:1996ww, Barbieri:1997tu, Feldmann:2008ja, Kagan:2009bn, Barbieri:2012uh}. 
In this construction, the first two families form a $U(2)$ doublet, while the third family is a singlet. In the symmetric limit, only the third generation acquires mass, providing a decent starting point. Symmetry breaking typically occurs in two stages: first, the $U(2) \equiv SU(2) \times U(1)$ symmetry is broken to an intermediate $U(1)$, generating a small mass for the second generation; subsequently, the $U(1)$ symmetry is broken, yielding an even smaller mass for the first generation. For recent developments, see section~\ref{sec:U2}.

\bigskip
\textit{Radiative models} propose that lighter SM fermion masses arise from loop corrections rather than at the tree level~\cite{Weinberg:1972ws}. In these scenarios, mass hierarchies arise from the suppression associated with loop factors, which intriguingly coincide with the two orders of magnitude separating the masses of adjacent generations. The absence of tree level Yukawa couplings for the light generations can, for example, be imposed by using flavor symmetries.
The breaking of a flavor symmetry is then communicated from a separate sector of the model to the SM fermions via loops. Additional fields and interactions are introduced to facilitate the necessary loop diagrams. 
In contrast to FN models, their masses can be at the same scale as flavor breaking, as the role of $\varepsilon$ is effectively replaced by a loop factor.
The wide range of possible quantum numbers and interactions for the new fields gives rise to a plethora of models, both supersymmetric~\cite{Banks:1987iu, Arkani-Hamed:1996kxn, Borzumati:1999sp, 
Altmannshofer:2014qha} and non-supersymmetric~\cite{Barr:1979xt,  Balakrishna:1988ks, Dobrescu:2008sz, Graham:2009gr}.

\subsubsection{Geometry-Based Models} Geometry-based models utilize extra spatial dimensions to address flavor hierarchies. In these frameworks, the SM fields propagate in higher-dimensional spaces and are localized at different positions in the extra dimension. The overlap of fermion wavefunctions with the Higgs field determines the effective four-dimensional Yukawa couplings. This overlap can depend exponentially on the parameters of the model. Adjusting localization parameters, therefore, allows one to generate the observed mass hierarchies without the need for hierarchical fundamental couplings~\cite{Arkani-Hamed:1999ylh}.

\begin{figure}[tb] \centering
  \includegraphics[width=.6\textwidth]{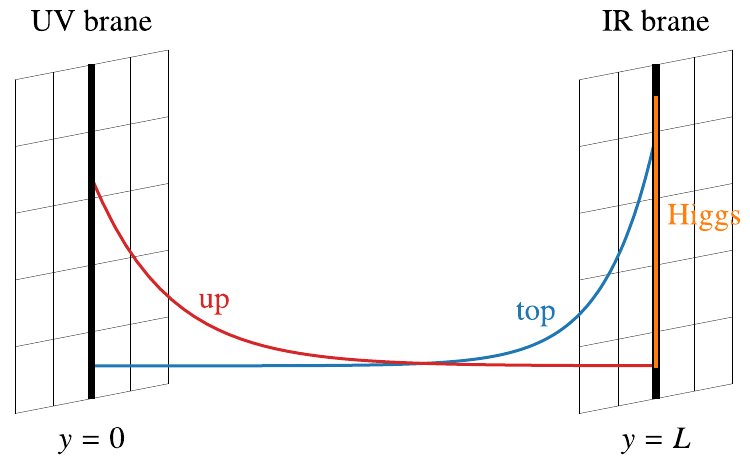}
  \caption{Illustration of the up quark and top quark extra-dimensional wave function profiles in a Randall-Sundrum model. The up quark is localized towards the UV brane, while the top towards the IR brane. The chosen bulk mass parameters are $c_u = 0.7$ and $c_t = 0.2$. The Higgs is shown localized on the IR brane. The size of the extra dimension is set to $L \simeq 37$ in Planck units, such that $e^{L} \simeq M_\text{Pl}/(1\,\text{TeV})$.}
  \label{fig:RS}
\end{figure}

A concrete example in which this mechanism can be at work is the Randall-Sundrum (RS) model~\cite{Randall:1999ee} that introduces a \textit{warped extra dimension} with a non-factorizable geometry, where the metric depends exponentially on the extra-dimensional coordinate. The extra dimension is bounded by an infrared (IR) brane and an ultraviolet (UV) brane. The warping leads to exponential factors that naturally generate hierarchies.
The Higgs naturalness problem is addressed if the Higgs resides on the IR brane. Fermions that propagate in the bulk have profiles that interpolate between the UV and IR branes and can be localized at different positions along the extra dimension, resulting in hierarchical Yukawa couplings with the Higgs. In such a case, the effective Yukawa couplings are given by 
\begin{equation}
\mathcal L \supset - \hat Y_u^{pr} f_{q_p} f_{u_r}   \, \overline{q}_p \widetilde{H} u_r -\hat Y_d^{pr} f_{q_p} f_{d_r} \, \overline{q}_p H d_r -\hat Y_e^{pr} f_{\ell_p} f_{e_r} \, \overline{\ell}_p H e_r\,,
\end{equation}
where the $\hat Y_{u,d,e}$ are the 5-dimensional Yukawa couplings that can be $\mathcal O(1)$, and the $f_\Psi$ are the wavefunctions evaluated on the IR brane. One finds
\begin{equation}
f_\Psi \simeq \begin{cases} \sqrt{1 - 2 c_\Psi} ~, \quad & \text{for} ~~ c_\psi < 1/2 ~, \\ \sqrt{2 c_\Psi - 1} ~ e^{-(c_\Psi - 1/2) L} ~, \quad & \text{for} ~~ c_\psi > 1/2 ~,\\ \end{cases}
\end{equation}
where $L$ is the size of the extra dimension and the $c_\Psi$ are called bulk masses that are free parameters. Heavier fermions, such as the top quark, are more localized towards the IR brane ($c_\Psi < 1/2$), resulting in $\mathcal O(1)$ Yukawa couplings, while light fermions are localized towards the UV brane ($c_\Psi > 1/2$) resulting in exponentially small Yukawas~\cite{Grossman:1999ra, Gherghetta:2000qt, Huber:2000ie}.
A typical example is shown in Figure~\ref{fig:RS}.

The flavor-dependent, extra-dimensional profiles of the SM fermions imply a rich flavor phenomenology~\cite{Agashe:2004cp, Blanke:2008zb}. In generic setups with warped extra dimensions, the constraints from meson mixing are very stringent and push the new physics scale to $\mathcal{O}(10)$\,TeV, thus creating an obstacle for a fully natural solution of the electroweak hierarchy problem. Extra dimensional models can be protected from large FCNCs by using bulk and brane flavor symmetries~\cite{Cacciapaglia:2007fw}. However, in those models, the flavor hierarchies are no longer generated by the attractive mechanism of wave function overlap.

\bigskip
\textit{Partial compositeness} proposes that the SM fermions are mixtures of elementary and composite states arising from a strongly coupled sector not far above the electroweak scale. Models of partial compositeness were originally introduced to tame FCNCs in technicolor models~\cite{Kaplan:1991dc}. The light fermions are mostly elementary, while the heavy top quark is mostly composite. Per the AdS/CFT correspondence, models of partial compositeness map onto models with warped extra dimensions, with the amount of compositeness given by the localization of the fermions along the extra dimension.

Composite Higgs models (see~\cite{Bellazzini:2014yua, Panico:2015jxa} for reviews) typically incorporate partial compositeness to generate fermion masses. The Higgs field emerges as a composite state from the strong sector, and the mixing between elementary fermions and composite operators leads to effective Yukawa couplings after electroweak symmetry breaking. This framework naturally explains the large top quark mass and addresses the hierarchy problem by protecting the Higgs mass from large radiative corrections. The phenomenology closely follows one of the RS models, and large FCNCs can be avoided using flavor symmetries~\cite{Csaki:2008zd, Redi:2011zi, Glioti:2024hye}.

\subsection{Model Building to Control New Physics Flavor Violation}

As discussed in section~\ref{sec:NP_puzzle}, generic BSM sources of flavor violation lead to prohibitively large contributions to FCNCs unless the NP scale is far above the electroweak scale. Instead, NP close to the electroweak scale can be viable if it exhibits a non-generic flavor structure that does not excessively violate the approximate flavor symmetries of the SM.

From a bottom-up point of view, an attractive but fairly restrictive option is the assumption that there are no new sources of flavor (and CP) violation beyond the SM Yukawa couplings. This is known as the {\it Minimal Flavor Violation} (MFV) hypothesis~\cite{Chivukula:1987py, Hall:1990ac, Buras:2000dm, DAmbrosio:2002vsn}. In an MFV scenario, NP contributions are controlled by the same hierarchical structure that suppresses FCNCs in the SM. To implement this idea systematically, the SM Yukawas are promoted to spurion fields with appropriate transformation properties under the flavor symmetry to formally restore flavor invariance of the Yukawa Lagrangian.
The SM flavor symmetry group is defined by the fermion kinetic terms
\begin{equation}
 G_F = U(3)_q \times U(3)_u \times U(3)_d \times U(3)_\ell \times U(3)_e ~,
\end{equation}
with the five $U(3)$ factors corresponding to three copies of the left-handed quark doublets, the right-handed up quarks, the right-handed down quarks, the left-handed lepton doublets, and the right-handed charged leptons, respectively. The Yukawa couplings are interpreted as bi-triplets of the relevant $SU(3)$ factors $Y_u = ({\bf 3}_q, \bar {\bf 3}_u)$, and so on.
A BSM setup is said to be minimal flavor violating if all new physics interactions are formally invariant under the SM flavor group. This implies that they are either completely flavor-blind or must be expressed as appropriate powers of Yukawa couplings. In addition, prohibiting new sources of CP violation is required to avoid stringent limits from EDMs. Extensions to the lepton sector are also possible~\cite{Cirigliano:2005ck}. 

In an MFV framework, constraints from meson mixing and other flavor observables are significantly eased, making new physics near the TeV scale a feasible possibility. Furthermore, MFV predicts correlations between $ b \to s $, $ b \to d $, and $ s \to d $ transitions, with the relative magnitudes of NP effects governed by the corresponding CKM factors. It is important to emphasize that MFV does not aim to resolve the flavor puzzle of the SM. Instead, it leverages the hierarchical flavor structure of the Yukawa couplings to constrain NP effects without providing an explanation for the origin of these hierarchies. 

\bigskip
Attractive alternatives to MFV that, in addition, can also partially address the SM flavor puzzle are models based on the {\it $U(2)^5$ flavor symmetry}. This symmetry (or its subgroups) could be responsible for the hierarchical pattern of fermion masses and mixings (see section~\ref{sec:U2} below) and, at the same time, suppress FCNCs in a sufficient way. The first and second generations of fermions $f_a$ ($a=1,2$) transform as a $ U(2)_f $ doublet, while the third generation is assigned as a $ U(2)_f $ singlet. The symmetric limit predicts that only the third generation acquires mass, leaving the first two generations massless. This serves as a plausible starting point for understanding the hierarchical structure of the Yukawa couplings. Furthermore, the $ U(2)^5 $ symmetry provides protection for flavor transitions between the first and second generations, which are subject to the most stringent experimental constraints. Given that $ y_t \sim 1 $, $ U(2)^5 $ provides a much better approximate symmetry than $U(3)^5 $. The realistic mass spectrum of the first and second generations, along with the CKM mixing, arises from a small breaking of $U(2)^5$, introducing small but testable deviations in flavor-changing processes.

Models based on the $U(2)$ flavor symmetries have been originally proposed in the context of SUSY \cite{Pomarol:1995xc, Barbieri:1995uv} and later formalized in a general context~\cite{Barbieri:2011ci, Fuentes-Martin:2019mun}. The framework of ``general MFV'' discussed in~\cite{Kagan:2009bn} shares many similar aspects. From a phenomenological point of view, $U(2)^5$ models are less restrictive than MFV models. While details depend on which spurions break the symmetry, one typically still finds strong correlations between $b \to s$ and $b \to d$ transitions. 

\bigskip
Yet another option to control NP effects in flavor changing processes is known as {\it flavor alignment}. The idea behind flavor alignment is that BSM interactions are approximately flavor diagonal, but not necessarily flavor universal, in the fermion mass eigenstate basis, i.e., in which the SM Yukawas are diagonal. This corresponds to an approximate $U(1)^6$ flavor symmetry in the quark sector, with each $U(1)$ factor acting on one of the six quark flavors. 
Most famously, in the context of supersymmetric models, the approximate alignment of quark Yukawa couplings and soft SUSY breaking squark masses to avoid large FCNCs has been proposed in~\cite{Nir:1993mx}. 

While flavor alignment suppresses flavor off-diagonal couplings, one can expect $\mathcal O(1)$ flavor non-universality. It is important to note that flavor alignment cannot be exact for new physics couplings of left-handed quark doublets unless the couplings are entirely flavor universal. Already in the SM, there is some misalignment between the left-handed up quarks and left-handed down quarks, as parameterized by the CKM matrix. A new non-universal flavor coupling can be aligned either with the left-handed up quarks or with the left-handed down quarks, but not with both of them simultaneously. An irreducible amount of flavor mixing of an order of the Cabibbo angle is predicted either for strange-to-down or charm-to-up transitions. Combining information from neutral kaon mixing and neutral $D$ meson mixing typically gives the strongest constraints on models with flavor alignment~\cite{Blum:2009sk}.

\section{RECENT DEVELOPMENTS}  \label{sec:recent}

In the previous section, we reviewed traditional approaches to addressing the SM and NP flavor puzzles. Now, we turn our discussion to a representative selection of recent developments in the field.

\subsection{Revisiting $U(2)$ and $SU(2)$ Symmetries} \label{sec:U2}

Models based on a \textit{single} approximate global $U(2)$ flavor symmetry~\cite{Pomarol:1995xc, Barbieri:1995uv, Barbieri:1996ww, Barbieri:1997tu} offer an elegant framework to explain flavor hierarchies by distinguishing the third family from the lighter two. While the main concept was outlined in section~\ref{sec:symmetry}, recent advancements have unveiled elegant realizations involving different charge assignments and symmetry-breaking patterns.

For concreteness, consider a $U(2)_q$ flavor symmetry acting on the left-handed quarks~\cite{Greljo:2023bix, Antusch:2023shi}. Imposing this symmetry leads to an extended symmetry of the Yukawa sector, $U(2)_q \times U(2)_u \times U(2)_d$~\cite{Feldmann:2008ja, Kagan:2009bn, Barbieri:2011ci}, where the latter two factors are accidental. To break $U(2)_q$, introduce two orthogonal spurion doublets, $ V = (0, a)^T $ and $ V' = (b, 0)^T $. This breaking structure generates quark Yukawa matrices $ Y_u $ and $ Y_d $, with entries in the first row of $ \mathcal{O}(b) $, second row of $ \mathcal{O}(a) $, and third row of $ \mathcal{O}(1) $.

Applying singular value decomposition, $ Y_f = L_f \hat{Y}_f R_f^\dagger $, the right-handed rotations $ R_{u,d} $ are matrices with $ \mathcal{O}(1) $ elements that transform $ Y_{u,d} $ into an upper triangular form. In contrast, the left-handed rotations $ L_{u,d} $ are perturbative matrices with mixing angles of $ \mathcal{O}(b/a) $, $ \mathcal{O}(a) $, and $ \mathcal{O}(b) $ for the $ 1$-$2 $, $ 2$-$3 $, and $ 1$-$3 $ sectors, respectively. The CKM mixing matrix $ V $ is then given by the product of left-handed rotations, $ V = L_u^\dagger L_d $. The singular values of both $ \hat{Y}_u $ and $ \hat{Y}_d $ are hierarchically ordered as $ \{b, a, 1\} $. By choosing parameters such that $ 1 \gg a \gg b \gg a^2 $, this broadly reproduces the observed hierarchies in quark masses between generations as well as the CKM mixing hierarchy. The biggest outlier to this picture is the bottom Yukawa coupling, which has a value of $y_b \sim 10^{-2}$.

This framework can be refined by also charging the right-handed up quarks $ u^c$, where $c$ denotes charge conjugation, resulting in a $ U(2)_{q+u^c}$ symmetry~\cite{Antusch:2023shi}. In this setup, the up-sector Yukawa matrix undergoes double suppression, where both left-handed and right-handed rotations are perturbative and of similar structure to $ L_{u,d} $ discussed earlier. Consequently, the singular values for the up-sector Yukawa matrix become $ \hat{Y}_u \sim \{b^2, a^2, 1\} $, while those for the down-sector remain $ \{b, a, 1\} $. A flavor-universal $ Z_2 $ symmetry is introduced whose breaking induces an overall downward shift in the down-quark mass spectrum by a factor of $\sim 10^{-2} $ (e.g., via a two-Higgs doublet model setup). This setup provides a better quantitative description and elegantly addresses the relative disparity between the up- and down-sector hierarchies, as the down-sector hierarchy is more compressed than the up-sector, and the third family is shifted downwards.

Leptons can naturally be included in both pictures. Since charged lepton mass hierarchies closely follow the down quarks, therefore, charging either left or right-handed leptons produces $ \hat Y_e \sim \{b, a, 1\} $. When left-handed leptons are charged under the flavor symmetry, the resulting perturbative left-handed rotations necessitate additional structure to account for the large observed PMNS mixing~\cite{Greljo:2023bix}. Alternatively, if the symmetry acts on right-handed charged leptons~\cite{Antusch:2023shi}, there are no constraints on the neutrino sector, naturally allowing for the large PMNS mixing observed in experiments. Therefore, the most promising symmetries are  $ U(2)_{q+e^{(c)}} $ and $ U(2)_{q+u^c+e^c} $. The latter, interestingly, is compatible with $SU(5)$ grand unification since ${\bf 10} = q \oplus u^c \oplus e^c$.

The IR features described above arise from specific UV dynamics. To illustrate one possible UV completion, consider gauging an $SU(2)$ flavor subgroup, which is anomaly-free in certain cases, such as $q+l$. The SM gauge-singlet flavon field, $\Phi$, is a flavor doublet whose VEV breaks the gauged flavor symmetry at
high energies (PeV scale and above) to comply with the bounds from cLFV~\cite{Greljo:2023bix, Darme:2023nsy}.
The light Yukawa couplings emerge from a dimension-5 operator in the symmetry-restored EFT, such as $\frac{1}{\Lambda} \bar{q} \Phi H d_p$. When $\langle \Phi \rangle = (0, v_\Phi)^T$, this operator generates the Yukawa coupling for one family, while insertions of $\tilde{\Phi} = \epsilon \Phi^*$ provide masses for the other family. With minimal field content completing the operator---for instance, a single copy of a vectorlike quark weak doublet and a flavor singlet---the two operators become proportional, leading to Yukawa matrices with rank two. Consequently, the minimal field content enforces an accidental $U(1)$ flavor symmetry for the first family at tree level, naturally establishing the $1$-$2$ hierarchy. A compelling mechanism to generate masses for the first family is through radiative corrections~\cite{Greljo:2023bix}. Rather than introducing new UV states, new IR states $S$ could run in the loop, with the first-family masses exhibiting only logarithmic sensitivity to their mass scale $m_S$, provided $m_S \lesssim m_\Psi$. This approach fits well with minimal quark-lepton unification, further enhancing the elegance of the mechanism~\cite{Greljo:2024zrj}.

This class of models, for sufficiently low flavor-breaking scales, predicts intriguing experimental signatures, including charged lepton flavor violation processes such as $\mu \to e$ conversion on heavy nuclei. By the end of this decade, upcoming experiments like Mu2e and COMET are expected to make significant advancements, potentially probing the flavor-breaking scale by another order of magnitude.

\subsection{Flavor Deconstruction}

Flavor deconstruction introduces a theoretical framework in which a gauge group $G$ is extended to $G^3$ in the UV, with one $G$ factor assigned to each fermion family. The fermion fields, $ f_i $ ($i=1,2,3$ for flavor), are each charged under their respective $ G_i $ factors, ensuring that anomaly cancellation occurs automatically within each family. The group $ G $ may represent the full SM gauge group, a subgroup of it, or an extension incorporating partial or complete quark-lepton unification. This paradigm, involving non-universal gauge symmetries, has gained momentum in recent years as a potential avenue for understanding flavor hierarchies as a consequence of multi-scale flavor dynamics, see for example~\cite{Davighi:2023iks, Greljo:2024ovt, Bordone:2017bld, Allwicher:2020esa, FernandezNavarro:2023rhv, Barbieri:2023qpf, Fuentes-Martin:2024fpx}. The spontaneous symmetry breaking unfolds through a two-step sequence. Initially, the extended gauge group $G_1 \times G_2$ breaks to a diagonal subgroup $G_{12}$, driven by a scalar field $\phi_1$ that acquires a vev, $\langle \phi_1 \rangle$. Subsequently, the residual $G_{12} \times G_3$ symmetry breaks to the universal gauge group $G$ via another scalar field $\phi_2$, which links $G_2$ and $G_3$. 

The spontaneous symmetry breaking to diagonal subgroups is naturally achieved using scalar link fields in nontrivial representations under both gauge factors. To economize the model, the fields are typically chosen so that $ \phi_1 f_1 $ and $ f_2 $ share the same gauge representation, as do $\phi_2 f_2$ and $f_3$.  The SM Higgs field $H$, in contrast, is charged only under $G_3$; therefore, the small Yukawas are generated via the link fields $ \phi_{2}$ and $ \phi_{1}$. This configuration offers a compelling mechanism for generating flavor hierarchies: the inter-family mass ratios are dictated by the suppression factors $ \epsilon_1 = \langle \phi_1 \rangle / \Lambda_1 $ and $ \epsilon_2 = \langle \phi_2 \rangle / \Lambda_2 $, where $\epsilon_i \sim \mathcal{O}(10^{-2}) $. $\Lambda_{i}$ denotes the scale behind the effective operators responsible for the low-energy Yukawa couplings.

This hierarchical structure, with $ \langle \phi_1 \rangle \gg \langle \phi_2 \rangle > \langle H \rangle $, effectively separates the dynamics of the light and heavy families. The Higgs field is predominantly exposed to the third family dynamics~\cite{Greljo:2018tuh, Allanach:2018lvl, Fuentes-Martin:2020bnh, Covone:2024elw} while being effectively shielded from the higher energy scales associated with the lighter families. This framework allows for the TeV-scale NP with built-in protection against FCNCs. Moreover, the multiscale structure preserves the stability of the electroweak scale, ensuring the finite naturalness of the Higgs mass~\cite{Davighi:2023iks}. In essence, this setup elegantly connects the flavor puzzles of both the SM and NP to the electroweak hierarchy problem. The IR phenomenology is governed by the $U(2)^5$ flavor symmetry. These models predict large effects in the third generation, specifically $B$-hadron and $\tau$ decays. Direct LHC searches for new resonances coupling mostly to the third generation are less constraining than those for the flavor-universal $U(3)^5$ NP due to suppressed production from valence parton distributions and difficulties detecting the third generation in the final state~\cite{Davighi:2023evx, Davighi:2023xqn, Capdevila:2024gki, FernandezNavarro:2024hnv}. This is most clearly exemplified by the constraints on semileptonic interactions derived from high-mass Drell-Yan processes~\cite{Greljo:2017vvb}.

Furthermore, flavor deconstruction frameworks can be embedded within a deeper UV context, such as warped extra dimensions~\cite{Fuentes-Martin:2022xnb}, enriching their theoretical appeal. It has also been proposed as a step toward electroweak flavor unification~\cite{Davighi:2022fer} or as a framework capable of producing intriguing three-peaked stochastic gravitational wave signatures~\cite{Greljo:2019xan}. These would arise from a sequence of cosmological phase transitions in the early universe, with the ratios of peak frequencies encoding the observed flavor hierarchies. Moreover, the intrinsic multiscale structure that gives rise to quark and charged lepton hierarchies can still accommodate the large (and seemingly anarchic) neutrino mixing in specific constructions~\cite{Greljo:2024ovt}.

\subsection{Revisiting $U(1)$ Symmetries}

As discussed in section~\ref{sec:symmetry}, Froggatt-Nielsen models address the flavor problem by assigning flavor-dependent charges to the SM fermions under a $U(1)$ flavor symmetry. This mechanism ensures that masses and mixing angles are parametrically determined by a charge-dependent power of a small spurion, $\epsilon^n$. A notable example of such a framework was proposed long ago~\cite{Leurer:1993gy}, using $\epsilon \simeq \lambda \simeq 0.2$ to reproduce the observed hierarchical masses and CKM mixing angles successfully. However, it is important to emphasize that the choice of charge assignments is not unique. A wide range of viable configurations exists, as free $\mathcal O(1)$ parameters can be tuned to accommodate observations with different sets of charges or alternative values of the expansion parameter $\epsilon$. As is often the case, there is a trade-off between the simplicity of charge assignments and the degree of variation allowed in the $\mathcal O(1)$ parameters. In~\cite{Fedele:2020fvh, Cornella:2023zme}, numerous viable charge assignments have been identified, including some with remarkably minimal charges. Analogously, in the framework of discrete FN models based on the $\mathbb{Z}_N$ group, the simplest realizations occur for small values of $ N \ge 4$~\cite{Greljo:2024evt}. These assignments produce distinct deviations in flavor-changing processes, enabling future precision measurements to distinguish between different classes. 

NP scenarios that account for the hierarchical flavor patterns of SM fermions often leave distinct imprints on physics beyond the SM. This is well established in FN models, where the horizontal $ U(1) $ symmetry dictates the flavor structure of NP couplings. Deviations from the expected scaling may arise due to additional UV dynamics and can be systematically classified and parameterized by higher powers of the flavor-breaking spurion~\cite{Asadi:2023ucx}. While these extensions enhance the flexibility of FN models in fitting low-energy flavor data, their size is constrained by self-consistency conditions.

\bigskip
Beyond their prominent role in FN models, flavor-dependent $ U(1) $ symmetries are significant in other contexts as well. For instance, the accidental symmetries of the SM---the three lepton family numbers $ L_e $, $ L_\mu $, $ L_\tau $, and baryon number---are observed to be exact (perturbative) symmetries for vanishing neutrino masses. This could be because they arise as global remnants of a spontaneously broken gauge symmetry. While $ B+L $ is anomalous, many linear combinations of these symmetries can be consistently gauged if the SM fermion content is extended by three right-handed neutrinos~\cite{Altmannshofer:2019xda, Greljo:2021npi}. 

A notable example is the difference between lepton flavor numbers, such as $L_\mu - L_\tau$, which remains anomaly-free even without right-handed neutrinos. A comprehensive catalog of anomaly-free lepton-non-universal $ U(1) $ charges in the SM extended with $ +3\nu_R $ was developed in~\cite{Greljo:2021npi}, building on the general classification framework of~\cite{Allanach:2018vjg}. In this scenario, lepton flavor universality is generically violated, while cLFV is strongly suppressed~\cite{Altmannshofer:2014cfa, Alonso:2015sja}. These symmetries offer a compelling approach to (partially) address the NP flavor puzzle, particularly by explaining the absence of cLFV, thereby permitting a lower NP scale. Furthermore, certain chiral charge assignments enable radiative mass generation, offering a partial solution to the SM flavor puzzle~\cite{Greljo:2021npi}. As an additional application, a class of lepton-flavored gauged $U(1)$ models predicts exact proton stability, ensured by an unbroken discrete remnant symmetry~\cite{Davighi:2022qgb}.

In phenomenological studies, lepton flavor universality violation in $B$-meson decays naturally emerges in these models, either through a massive $Z^\prime$ gauge boson associated with the spontaneously broken $U(1)$ symmetry or via leptoquarks charged under $U(1)$~\cite{Altmannshofer:2019xda, Davighi:2020qqa, Greljo:2021npi}. Additionally, a $ Z^\prime $ boson with a mass in the range of $ 10\text{–}100 \,\text{MeV} $ could significantly influence the anomalous magnetic moment of the muon~\cite{Greljo:2022dwn}.

\subsection{Flavor symmetries in the SMEFT} \label{sec:SMEFT}

As already discussed in section~\ref{sec:NP_puzzle}, the SMEFT provides a valuable theoretical framework to describe the low-energy effects of a generic high-energy theory beyond the SM. The SMEFT Lagrangian consists of an infinite series of higher-dimensional local operators built from SM fields and symmetries, with the details of the underlying theory encoded in the Wilson coefficients. Investigating the correlations and patterns in these Wilson coefficients helps map out the landscape of possible short-distance physics beyond the SM.

However, this broad approach comes with challenges. One of the main difficulties in using SMEFT is the dramatic increase in independent Wilson coefficients, largely due to the complexity introduced by flavor. For instance, when considering baryon-number conserving dimension-six operators at leading order, the number of parameters surges from 59 with a single generation to a staggering 2499 for three generations~\cite{Alonso:2013hga}. Even in the renormalizable SM, flavor-related parameters dominate---the Yukawa sector alone introduces 13 parameters, compared to three in the gauge sector and two in the scalar potential. Understanding the flavor structure is a critical open issue in the SMEFT.

Global flavor symmetries such as $U(3)^5$ and $U(2)^5$ and their breaking patterns offer an effective way to structure the SMEFT, grouping theories beyond the SM into universality classes~\cite{Faroughy:2020ina, Greljo:2022cah}. Instead of considering the resulting model dependence as a limitation, it should be seen as a valuable opportunity to gain systematic insights into UV physics through experimental data.

Initial work has been done on developing a theoretical framework to classify flavor symmetries relevant to SMEFT. For example, in~\cite{Greljo:2022cah}, the flavor structure of lepton and baryon number-conserving dimension-6 operators within the SMEFT was investigated. Several well-motivated flavor symmetries and symmetry-breaking patterns were proposed as competing hypotheses regarding the ultraviolet dynamics beyond the SM. For each scenario, independent operators were explicitly constructed and counted up to the first few orders in the spurion expansion (see Fig.~1 in~\cite{Greljo:2022cah}  for the leading order), offering setups ready for phenomenological studies and global fits. 
Understanding the flavor structure of the SMEFT renormalization group is essential~\cite{Machado:2022ozb}, paving the way for phenomenological studies such as those in~\cite{Allwicher:2023shc, Grunwald:2023nli, Greljo:2023bdy}.

\subsection{Flavor Symmetries and Axions} \label{sec:axions}

As discussed in section~\ref{sec:symmetry}, the SM flavor puzzle can be addressed by a spontaneously broken horizontal $U(1)$ flavor symmetry {\`a} la Froggatt-Nielsen. Such a dynamical structure implies observable consequences. In particular, if the $U(1)$ symmetry is not gauged, its spontaneous breaking gives a pseudo-Nambu-Goldstone boson, which is naturally light and can be searched for in the experiment. Furthermore, if the $U(1)$ flavor symmetry is anomalous under QCD, the Goldstone can act as an axion and solve the strong CP problem~\cite{Wilczek:1982rv}. In this context, the pseudo-Nambu–Goldstone boson is sometimes referred to as an ``axiflavon''~\cite{Calibbi:2016hwq} or ``flaxion''~\cite{Ema:2016ops}, see also \cite{Arias-Aragon:2017eww, Bjorkeroth:2018ipq, Bonnefoy:2019lsn, DiLuzio:2023ndz}. 
Generation-dependent $U(1)$ charges will, in general, give rise to flavor-changing couplings of the axion to fermions. The couplings are proportional to the difference in $U(1)$ charges and suppressed by the axion decay constant~$f_a$.

In general, axion-like particles (ALPs) are well-motivated and naturally light extensions of the SM, even if they do not address the strong CP problem. An example of an ALP that is connected to a solution of the SM flavor puzzle is given in~\cite{Greljo:2024evt}. The considered ALP is part of Froggatt-Nielsen models that are based on discrete $\mathbb{Z}_N$ symmetries. In such models, the ALP mass can be considerably heavier compared to continuous $U(1)$ models or the QCD axion. The $\mathbb{Z}_4$ and $\mathbb{Z}_8$ symmetries emerge as particularly promising examples.

An axion or ALP with flavor-changing couplings can give observable effects in a broad range of flavor observables, including rare decays of mesons, neutral meson oscillations, rare lepton decays, and electric and magnetic dipole moments of quarks and leptons~\cite{Choi:2017gpf, Bjorkeroth:2018dzu, MartinCamalich:2020dfe, Bauer:2020jbp, Carmona:2021seb, Bauer:2021mvw}. Of particular interest are two-body decays of mesons and leptons that can be parametrically enhanced compared to the standard weak decays of these particles. For example, in the presence of a flavor changing strange-down coupling of an axion, the branching ratio of the decay $K \to \pi a$ scales as $\text{BR}(K \to \pi a) \propto m_W^4/(f_a m_K)^2$. An analogous enhancement is present in $B \to K a$ or $\mu \to e a$. If the axion is light, $m_a \lesssim 1$\,MeV, its lifetime is macroscopic, and it will give a missing energy signature in detectors. Relevant searches are therefore, for example, $K \to \pi + \text{invisible}$, $B \to K + \text{invisible}$, and $\mu \to e + \text{invisible}$~\cite{Calibbi:2020jvd, Jho:2022snj, Knapen:2023zgi}.
Constraints from existing searches for $K \to \pi + \text{invisible}$ and $\mu \to e + \text{invisible}$ put bounds on the axion decay constant of around $f_a \gtrsim 10^{12}$\, GeV~\cite{MartinCamalich:2020dfe} and $f_a \gtrsim 10^9$\, GeV~\cite{Calibbi:2020jvd}, respectively, if the couplings are flavor anarchic. Heavier ALPs can give visible decay signatures, for example, $a \to \gamma \gamma$ or $a \to e^+ e^-$, that can be either prompt or displaced and that can be effectively searched for at flavor factories.

Bounds derived from flavor observables are often competitive with those from collider and beam dump experiments and astrophysical observations covering complementary parameter space regions. They are particularly significant in the MeV to GeV mass range, where other probes typically provide weaker constraints.

\subsection{Flavor Clockwork Models}

The clockwork mechanism~\cite{Giudice:2016yja} is a fairly recent idea to generate exponentially small interactions in theories without small parameters at the fundamental level. A fermionic clockwork starts with $2N + 1$ chiral fermions that are chained together by mass parameters chosen such that a single chiral symmetry is shared among all the fermions. This gives $N$ massive Dirac fermions and a single massless chiral mode. The massless mode overlaps with the $N$th original fermion that is suppressed by a clockwork factor $1/q^N$. If the free parameter $q$ is larger than one, $q > 1$, exponentially small couplings of the zero mode are thus generated. The clockwork mechanism has been successfully applied to explain the hierarchical structure of the quark, lepton, and neutrino masses~\cite{Patel:2017pct, Ibarra:2017tju, Alonso:2018bcg, Hong:2019bki, AbreudeSouza:2019ixc, Kang:2020cxo, Babu:2020tnf}. In these scenarios, indirect constraints from low-energy flavor observables, such as rare decays and meson oscillations, can be effectively controlled. Furthermore, the new degrees of freedom predicted by the fermionic flavor clockwork may be within reach of collider experiments.

Flavor clockwork models share some similarities with FN setups. The role of the $U(1)$ charges is played by the length $N$ of the clockwork chains, while the symmetry-breaking spurion $\langle \phi \rangle / M \ll 1$ is replaced by the clockwork gear ratio $1/q$. In fact, the clockwork mechanism has motivated the construction of ``inverted'' Froggatt-Nielsen models, which have the expansion parameter $M/\langle \phi \rangle$~\cite{Smolkovic:2019jow}. Interestingly, such constructions are anomaly-free, and the horizontal $U(1)$ symmetries can be gauged. In principle, the corresponding $Z^\prime$ gauge bosons can be light and can be searched for using precision flavor, astrophysics, and beam dump experiments.

Supersymmetric versions of flavor clockwork models have been constructed in~\cite{Altmannshofer:2021qwx}. The zero modes of the clockwork are identified with the fermions and sfermions of the minimal supersymmetric SM. In addition to generating a hierarchical fermion spectrum, the clockwork also predicts a specific flavor structure for the soft SUSY-breaking sfermion masses. Interestingly, the simplest setup predicts large flavor mixing among first and second-generation squarks, akin to the simplest supersymmetric $U(1)$ FN models. Constraints from kaon oscillations require the masses of either squarks or gluinos to be at the PeV scale or above~\cite{Altmannshofer:2013lfa}.

\subsection{Modular and Eclectic Flavor Symmetries}

Modular flavor symmetries were originally considered in the context of neutrino mass model building~\cite{Feruglio:2017spp, Ding:2023htn}. While discrete flavor symmetries can successfully describe the mixing pattern in the neutrino sector at zeroth order, it is challenging to obtain neutrino masses and mixings that precisely match the observed values in minimal models. This challenge has motivated research into a class of supersymmetric models that generalize discrete symmetries using so-called modular forms.\footnote{Interestingly, the holomorphicity of the superpotential is an important ingredient to ensure consistent modular transformation properties. Exploration of non-supersymmetric models with modular invariance has only just started~\cite{Ding:2020zxw, Qu:2024rns, Ding:2024inn}.} 

In practice, Yukawa couplings are promoted to modular forms $Y(\tau)$ that depend on the complex modulus $\tau$. If the vev of the modulus is close to a symmetry point~\cite{Okada:2020ukr, Feruglio:2021dte, Novichkov:2021evw}, the Yukawa couplings can show a hierarchical pattern that is determined by the ``weight'' and ``level'' of the modular form. In contrast to Froggatt-Nielsen models, the achievable hierarchy patterns are more constrained, enhancing their predictive power. These are determined by finite modular groups, which, in the simplest cases, are isomorphic to permutation groups.

Models based on modular flavor symmetries usually have a much smaller field content than FN models, which contain one or more flavon fields for spontaneous flavor breaking, as well as a fairly large sector of vector-like matter that mediates the flavor breaking to the SM fermions. Simple models based on modular flavor symmetries do not require such ingredients. Interestingly, the most economical modular flavor model predicts the mass splittings of neutrinos and the three mixing angles in the lepton sector in terms of a single complex parameter, the vev of the modulus $\tau$~\cite{Feruglio:2017spp}. Despite the challenge of reproducing five known observables using only one complex parameter, this model performs relatively well.

To obtain a hierarchical quark and charged lepton spectrum, as well as a hierarchical CKM matrix more readily, one can add a ``weighton'' field, which provides suppression of Yukawa couplings according to the chosen modular weights of the SM fermions~\cite{King:2020qaj}. To get a full agreement with observed masses and mixing angles, one typically needs to add free parameters by writing the Yukawa couplings as linear combinations of more than one modular form. Additional free parameters can also come from the K{\"a}hler potential~\cite{Chen:2019ewa}. Models that achieve fully realistic fermion masses and mixing can be found, for example, in~\cite{Lu:2019vgm, Okada:2019uoy, Liu:2020akv}.

Modular flavor symmetries can also directly be combined with a $U(1)$ flavor symmetry, which is responsible for hierarchies among the three generations of quarks and lepton masses and CKM angles. Such a setup has been constructed, for example, in~\cite{deAnda:2018ecu} in the context of a $SU(5)$ grand unified theory.  In another recent development involving modular symmetries, an intriguing connection between the flavor puzzle and the strong CP problem has been proposed~\cite{Feruglio:2024ytl}.

Further developments in this direction of flavor model building are eclectic flavor symmetries~\cite{Baur:2019kwi, Nilles:2020nnc} that combine in a nontrivial way modular flavor symmetries, ``traditional'' discrete flavor symmetries, and CP-like symmetries in top-down string motivated models.

\subsection{Flavor from Higher-Form and Non-Invertible Symmetries}

Generalized symmetries extend the traditional notion of symmetries~\cite{Gaiotto:2014kfa}. 
So-called higher-form global symmetries act not just on point-like objects like particles but on higher-dimensional structures, such as lines, surfaces, or volumes. Non-invertible global symmetries are permitted to have a product operation that goes beyond that of a group. In particular, the inverse under the product operation does not need to exist. 

Higher-form global symmetries and non-invertible global symmetries retain many features and properties of ``ordinary'' global symmetries, and they lead, for example, to selection rules on amplitudes. While the implications of these generalized symmetries are still under active investigation, they provide new powerful tools to analyze the structure of quantum field theories. They are also being applied to the study of particle physics phenomenology. For a recent review on the topic of generalized symmetries in particle physics, see~\cite{Brennan:2023mmt}.

Of particular interest to flavor model building is the result that non-invertible chiral symmetries can naturally lead to exponential hierarchies. As shown in~\cite{Cordova:2022ieu}, symmetries that suffer from abelian Adler-Bell-Jackiw anomalies can survive as non-invertible chiral symmetries. These non-invertible symmetries imply selection rules for couplings that can be broken by non-perturbative effects and give technically natural exponential hierarchies. The authors of~\cite{Cordova:2022ieu} apply this idea to generate exponentially small fermion masses and speculate that this mechanism may find applications in particle physics model building.

In the related work~\cite{Cordova:2022fhg}, the authors show that gauging the familiar $L_\mu - L_\tau$ symmetry can result in non-invertible symmetries that protect neutrino masses. Non-zero (but exponentially suppressed) neutrino masses can then arise non-perturbatively from instantons. It needs to be seen if such constructions can provide a fully realistic neutrino spectrum, including neutrino mixing, and if such a setup could be extended to the quark sector as well.

In~\cite{Kobayashi:2024yqq, Kobayashi:2024cvp}, the authors show how certain string-motivated non-invertible symmetries can be used to arrive at various textures of Yukawa matrices, including many examples with so-called texture zeros both in the quark and lepton sector, which can be good starting points for further flavor model building. 

Finally, Ref.~\cite{Cordova:2022qtz} initiates the study of the generalized global flavor symmetries of the SM and their connection with models of gauge unification. They observe that the ordinary SM flavor symmetries are intertwined with a one-form magnetic hypercharge symmetry, forming a so-called two-group structure. They identify all possible vertical unification patterns that are compatible with this structure. It might be interesting to extend such a study to horizontal symmetries and gauge-flavor unification.

\section{CONCLUSIONS}  \label{sec:conclusions}

The origin of flavor remains one of the key open questions in particle physics, driving efforts to extend the SM. This challenge is tied to two interconnected puzzles. First, the observed hierarchical patterns of SM fermion masses and mixing angles are highly non-generic, collectively referred to as the SM flavor puzzle, suggesting the need for an explanation beyond the SM. Second, NP near the electroweak scale typically induces unacceptably large contributions to flavor-changing transitions, introducing the so-called NP flavor puzzle. This puzzle suggests that NP must possess a non-generic flavor structure, particularly in scenarios addressing the Higgs naturalness problem, which often predicts new states around the TeV scale with significant couplings to SM particles. Flavor model building seeks to address both puzzles by constructing mechanisms to generate hierarchical flavor couplings, integrating them into NP models, and deriving testable phenomenological predictions.

In this review, we have provided a concise summary of the two major flavor puzzles in Section~\ref{sec:puzzles} and discussed traditional solutions in Section~\ref{sec:solutions}, including approaches that use flavor symmetries to generate hierarchies, as well as geometric approaches. Recent developments are reviewed in section~\ref{sec:recent}. 

Flavor models that are based on the popular $U(2)$ or $SU(2)$ flavor symmetries, as well as $U(1)$ models \`a la Froggatt-Nielsen (FN), continue to evolve and undergo refinement. Flavor models with a spontaneously broken global $U(1)$ symmetry predict the existence of light pseudo-Nambu-Goldstone bosons, such as axions or axion-like particles, characterized by a distinctive pattern of flavor-violating couplings. Developing innovative search strategies for these light, weakly coupled particles remains a highly active area of research.

The concept of flavor deconstruction has recently garnered significant interest. This framework proposes that gauge interactions are non-universal in the UV, with the SM flavor hierarchies emerging from hierarchical scales associated with each family. This setup also addresses the NP flavor puzzle, allowing for physics at the TeV scale, as anticipated by the Higgs hierarchy problem.

Flavor symmetries have been systematically applied to the SMEFT, offering a robust bottom-up framework to explore flavorful new physics. This approach enables a structured investigation of how flavor dynamics can manifest in extensions of the SM. Among recent developments, flavor models based on the clockwork mechanism proposed as an alternative to FN models generate hierarchies through a structured chain of interactions. Additionally, modular flavor symmetries have emerged as a novel tool in model building, particularly in the neutrino sector, where they offer elegant explanations for the observed flavor patterns by linking flavor symmetries to the geometry of extra dimensions. Finally, innovative concepts such as higher-form symmetries and non-invertible symmetries are beginning to shape the landscape of flavor model building. 

As precision flavor experiments push the boundaries of our understanding, flavor model building remains indispensable, serving as both a guide for interpreting new data and a cornerstone for uncovering the deeper principles that govern the structure of matter and the fundamental forces of nature.

\section*{DISCLOSURE STATEMENT}
The authors a.re not aware of any affiliations, memberships, funding, or financial holdings that might be perceived as affecting the objectivity of this review.

\section*{ACKNOWLEDGMENTS}
We thank Joe Davighi, Xavier Ponce Díaz, and Miguel Levy for their useful comments.
The research of WA is supported by the U.S. Department of Energy grant number DE-SC0010107. AG has received funding from the Swiss National Science Foundation (SNF) through the Eccellenza Professorial Fellowship ``Flavor Physics at the High Energy Frontier,'' project number 186866.

\bibliographystyle{ar-style5.bst}
\bibliography{bibliography}

\end{document}